\documentclass[10pt,superscriptaddress,twocolumn,floatfix,showpacs]{revtex4-2}
\usepackage{graphicx}
\usepackage{amsmath}
\usepackage{amsfonts}
\usepackage{amssymb}
\usepackage{epsfig}
\usepackage{epstopdf}
\DeclareGraphicsExtensions{.pdf,.eps,.png,.jpg,.mps}
\usepackage[pdftex]{color}
\usepackage{amsmath,graphicx,amssymb,braket,xcolor,subfigure,upgreek}
\usepackage[colorlinks, linkcolor=blue, citecolor=blue, urlcolor=blue, breaklinks=true]{hyperref}
\usepackage{microtype}
\usepackage{bbm}
\usepackage{color}
\usepackage{braket}

\usepackage{amsmath,amssymb,amsfonts,latexsym,color,dcolumn,bm,amsbsy}

\usepackage[english]{babel}
\usepackage{graphicx}
\usepackage[utf8]{inputenc}
\usepackage{textcomp}

\usepackage{braket}
\newcommand{\non}{\nonumber}
\begin{document}
	\title{Tunable Electromagnetically Induced Multi-Transparencies in Hybrid Optomechanical system Incorporating Atomic Medium}
	\author{M. Hunza}
	\affiliation{Department of Physics, COMSATS University Islamabad, 45550, Pakistan,}
	\author{M. Asjad}
	\email{asjad\_@yahoo.com}
	\affiliation{Department of Applied Mathematics and Sciences, Khalifa University, 127788, Abu Dhabi, UAE,}
	\author{T. Abbas}
	\affiliation{Department of Physics, COMSATS University Islamabad, 45550, Pakistan,}
	\author{M. Qasymeh}
	\affiliation{Electrical and Computer Engineering Department Abu Dhabi University, 5991 Abu Dhabi, UAE,}
	\author{B. Teklu}
	\affiliation{Department of Applied Mathematics and Sciences, Khalifa University, 127788, Abu Dhabi, UAE,}
	\affiliation{Center for Cyber-Physical Systems (C2PS), Khalifa University, 127788, Abu Dhabi, UAE,}
	\author{H. Eleuch}
	\affiliation{Department of Applied Physics and Astronomy, University of Sharjah, Sharjah, UAE,} 
	\affiliation{Institute for Quantum Science and Engineering,Texas AM University, College Station, TX 77843 USA.}
	\begin{abstract}
We consider a hybrid atom-optomechanical system incorporating N identical $\Lambda$-type atoms. The system is subjected to dual optical and phononic drives. We show that by exploiting the optomechanical linear and quadratic interactions, multiple electromagnetic transparency windows are attained. Furthermore, owing to the incorporated mechanical pump, the transparency windows are controlled and tuned. For instance, by adjusting the phase of the external mechanical pump,  additional controlling parameters are enabled, and the absorption/emission profiles are enhanced. Our present study provides an efficient approach to modifying propagating signals inside the quantum devices incorporating cavity-optomechanical systems.	
\end{abstract}
\date{\today}
\maketitle
\section{Introduction}
Since the achievement of electromagnetically induced transparency (EIT), several significant advancements in optical physics have been accessible \cite{Harris,zubairy}.  EIT is now an active area of research in both theoretical and experimental domains. This phenomenon has ushered into different applications such as the propagation of ultraslow light pulses \cite{hau1999, scully1999, Boutabba}, cooling of  ground-state atoms \cite{maier2016}, enhancement in the efficiency of nonlinear conversion \cite{Hakuta1993, Hemmer1995} etc. Further progress has also been made in the context of coherent control of a massive mechanical oscillator, leading to optomechanical induced transparency (OMIT) \cite{Agarwal2010, R2010, Safavi2011} as well as to optomechanical induced absorption (OMIA) \cite{singh2014,Agarwal2013,zhou2012}. OMIA is analogous to the EIT phenomenon whichhas been extensively studied in the presence of atomic media \cite{Wu2007,X2011,X2010}. Devices based on nano and micro mechanical systems can be coupled to many systems through magnetic coupling \cite{Rabl2009}, optical dipoleforce \cite{Michael2007,xiong2008} and by the radiation pressure caused by the optical field \cite{Sete2015, Eleuch2012, Juuso, Aspelmeyer14, qoptm,Eleuch2015} that paves the way for configurability in quantum systems.\\

The generic utility in optomechanical induced transparency (OMIT) is the interaction of  the optical mode with the mechanical mode that is noticeable in different experimental as well as theoretical  developments \cite{chang2011,han2011}. OMIT dependent phenomena like, nonlinear quantum domain \cite{kronwald2013, Borkje2013}, second and higher order side-bands \cite{H2012,H2016} include linear coupling. Likewise, configurations like membrane placed in-between the cavity  \cite{Sankey2010, asjad2013, asjad2014}, ultracold atoms \cite{Murch2008, Purdy2010}, have  been realized under quadratic optomechanical coupling (QOC). Furthermore, quadratic coupling of  the mechanical oscillator with the cavity field reveals the phenomena of slow light and two-phonon OMIT  \cite{Karuza2013, Zhan2013}. Such coupled systems can be used to measure displacements with  optimum precision in the field of quantum metrology \cite{Rugar2004}, quantum probes \cite{Candeloro} as well as the detection of  gravitational waves \cite{Braginsky2002}, engineering the micro-macro sytem entanglement \cite{Vitali07,Korppi18,asjad15, asjad16, Berihu} with important implications for quantum logic gates based on EIT schemes \cite{Feizpour}. 

Motivated from all of the above stated work, we investigate the behavior of OMIT, where an atomic ensemble is present, along with linear as well as quadratic coupling that are interacting  in the simultaneous presence of the strong optical field as well as a phonon pump of the external mechanical mode with the cavity mode field. Multiple tunable transparency windows are significantly observed in the presence of atomic media in cavity optomechanical systems. 

The paper is organized as follows. In Section \ref{sec2}, the details of the theoretical model at hand is discussed along with the Langvein equations whereas Section \ref{sec3}, contains the graphical representations of the analytical solution of the optical response of the weak field. Section \ref{sec4}, contains the conclusion of the results.
\section{model and equations} \label{sec2}
\begin{figure}[h]
\centering
\includegraphics[scale=0.40]{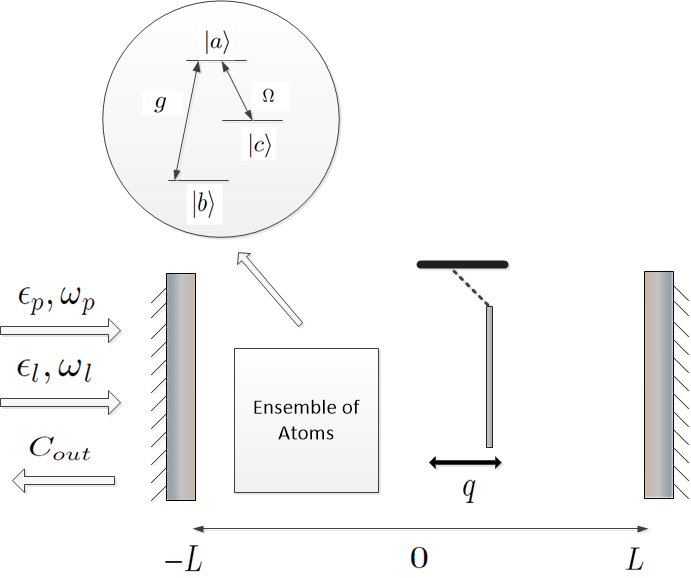}
\caption{Schematic of the system: Hybrid atom-optomechanical system with a membrane that is coupled linearly and quadratically to the cavity modes. The cavity is driven by strong laser field with amplitude $\Omega$ and frequency $\omega_{l}$ and a weak laser beam of amplitude $\tilde{\epsilon_ m}$ and  frequency $\omega_p$.}
\label{fig:cavityy}
\end{figure}

We consider an ensemble of N identical $\Lambda$-type three-level atoms confined
inside an optical cavity, that has an oscillating membrane placed between the optical cavity as shown in Fig. \ref{fig:cavityy}. Here, we are
not considering the membrane to be present in the equilibrium position, but rather, a more generalized
scenario where the membrane is oscillating and is coupled to both the linear and quadratic optomechanical coupling
via modes of the cavity. For the $i$th atom, a classical control  field with amplitude $\Omega$ and frequency $\nu$ induces transition between
levels $|a \rangle\leftrightarrow|c\rangle$ whereas, the cavity field with frequency $\omega_{o}$ interacts with atoms making a transition between the atomic levels $|a \rangle\leftrightarrow|b\rangle$. In addition to this, we have an external mechanical mode that induces a phononic pump that is also responsible for the vibrations in the
membrane.
The schematics are presented in Fig. ( \ref{fig:cavityy}).
Considering $\hbar= 1$ , the Hamiltonian of the  system is given by, 
\begin{multline}
\label{eq:ham}
H = H_C + H_A + H_M+ H_{MC} + H_{AC} + H_{CL}.
\end{multline}
The first term in the above Hamiltonian represents free energy of the cavity field which is given as,
\begin{eqnarray}\label{eq:ham1}
H_{C} &=& \omega_{o} c^{\dagger} c.
\end{eqnarray}
Here, ${c}^{\dagger}$ and ${c}$ design respectively the creation and annihilation operators of the field.
The second term in Eq. (\ref{eq:ham}) represents free energy of an ensemble of N identical $\Lambda$\ - type atoms which can be written as,
\begin{eqnarray}\label{eq:ham2}
H_A &=& \sum_{i=1}^{n} (\omega_{a} \sigma_{aa}^{i} +\omega_{b}\ \sigma_{bb}^{i}+\omega_{c}\ \sigma_{cc}^{i}).
\end{eqnarray}
The Bohr frequencies $\omega_{a}$, $\omega_{b}$ and $\omega_{c}$ relate to the atomic level $\vert a \rangle$, $\vert b \rangle$ and $\vert c \rangle$ respectively, together with the respective atomic operators  $\sigma_{aa}$, $\sigma_{bb}$ and $\sigma_{cc}$ that correspond to these atomic levels. The energy for the mechanical motion of the oscillating mirror is represented by $H_m$ in Eq. (\ref{eq:ham}) is given by,
\begin{equation}
H_M= \dfrac{\omega_m}{2}  (p^2+q^2).
\label{eq:ham3}
\end{equation}Here, $\omega_{m}$ is the frequency, $p$ is the momentum and $q$ is the displacement of the oscillating mirror. The coupling between cavity field and oscillating mirror represented by $H_{CM}$ in Eq. (\ref{eq:ham}) is given by,
\begin{equation}\label{eq:ham4}
H_{CM}=g_{1} q c^{\dagger}c+g_{2}q^{2}c^{\dagger}c.
\end{equation}The linear and the quadratic coupling constants between cavity field and oscillating mirror are represented here as, $g_1$ and $g_2$ respectively. These constants are defined as $g_1 = \frac{\partial \omega_c}{\partial q} x_o$ and $g_2 = \frac{\partial^2 \omega_c}{\partial q^2} \frac{x_o^2}{2}$. The zero point fluctuations in displacement of oscillating mirror are represented as $x_{o}$. The interaction of fields (both quantized and classical) with the atomic ensembles can be represented as,
\begin{equation}
H_{AC}=\sum_{i=1}^{N}(g c\sigma_{ab}^{i}+\Omega \exp(-i\nu t)\sigma_{ac}^{i}+ H.c).
\end{equation}
The interaction between atomic levels $|a \rangle\leftrightarrow|b\rangle$ with quantized field along with a coupling constant is shown as, $g = -\mu(\omega_{0}/ 2V\epsilon_{0})$. The electric dipole between two levels is represented as $\mu$ whereas, $V$ is the volume of the cavity. The atomic interaction between levels $|a \rangle\leftrightarrow|c\rangle$ are driven by classical field having the Rabi frequency $\Omega$. Furthermore the cavity is driven by a pump field and a stokes field with frequency $\omega_{l}$ and $\omega_{p}$ respectively.  The last term in Eq. (\ref{eq:ham}) accounts for strong optical pump, probe field and phonon driving and is given by
\begin{eqnarray}\label{eq:ham6}
H_{CL} =i c^{\dagger} (\epsilon_{l} e^{-i\omega_{l}t}+\epsilon_{p} e^{-i\omega_{p}t})-q\tilde{\epsilon}_m e^{i\delta t} - H.c, 
\end{eqnarray}
where $\epsilon_{l}=(2\kappa \wp_{l}/\omega_{l})^{1/2}$ is the amplitude of pump field whereas $\epsilon_{p}=(2\kappa\wp_{p}/\omega_{p})^{1/2}$ corresponds to the probe field's amplitude. $\wp_{l}(\wp_p)$ is the power of pump and probe field. The last term mentioned in Eq.(\ref{eq:ham6}) describes the phonon pump of the mechanical mode where, $\tilde{\epsilon}_m=\epsilon_{m}e^{i\Phi_m}$ is the strength of the phonon pump with phase $\Phi_m$ and $\delta=\omega_{p}-\omega_{l}$.  Now we define atomic ensemble operator $A=(1/N)^{1/2} \sum_{i=1}^{n} \sigma_{ba}^{i}$ and $C=(1/N)^{1/2} \sum_{i=1}^{n} \sigma_{bc}^{i}$ which satisfy the Bosonic relation $[C, C^\dagger]=1$, $[A, A^\dagger]=1$. 
Then the Heisenberg-Langevin equation for this system reads
	\begin{eqnarray}\label{eq:1}
		\dot{q}&=&\omega_m p,\nonumber\\
		\dot{p}&=& -\omega_m q-\gamma_m p-g_1 c^\dagger c-2g_2 c^\dagger c q+\tilde{\epsilon}_m\cos(\delta t)+\xi,\nonumber\\
		\dot{c}&=& -[\kappa + i(\omega_{0}+g_1 q+ g_2 q^{2})]c- i G_a  A +\epsilon_{l}e^{-i\omega_{l}t} +
		\epsilon_{p} e^{-i\omega_{p}t} \nonumber\\ &+&
	\sqrt{2\kappa} c_{in},\nonumber\\
		\dot{A}&=& -(i\omega_{ab}+\gamma_{1}) A-i G_a c-i\Omega e^{-i\upsilon t} C +\sqrt{2\gamma_{1}}A_{in},\nonumber\\
		\dot{C}&=& -(i\omega_{cb}+\gamma_{2})C - i\Omega e^{i\upsilon t} A +\sqrt{2\gamma_{2}} C_{in},
	\end{eqnarray}
where $G_a=gN^{1/2}$ is the effective atom filed coupling.
Here, the atomic decay rate between the three level atom is represented by $\gamma$. More precisely, $\gamma_{1}(\gamma_{2})$ is the decay rate of transition $|a \rangle\leftrightarrow|b \rangle(|a \rangle \leftrightarrow |c \rangle)$. Similarly the transition frequencies for $|a \rangle\leftrightarrow|b \rangle(|c \rangle \leftrightarrow |b \rangle)$ are represented as $\omega_{ab}=\omega_{a}-\omega_{b}$ and $\omega_{cb}=\omega_{c}-\omega_{b}$. The decay rate of the oscillating mirror is shown by $\gamma_{m}$ whereas, $\kappa$ is the decay rate of cavity field. 
Here, $\xi, A_{in}$, $C_{in}$ and $c_{in}$ represent the noise terms of mirror, atomic parts, and cavity respectively. The cavity field evolution is related to the oscillating mirror, that oscillates quadratically and is represented as $ q^2$ as shown in Eq. (\ref{eq:1}) which corresponds to the mechanical mode of oscillating mirror. Therefore we define $Q=\langle q^2\rangle$, $P=\langle p^2\rangle$ and $X=\langle p q+q p\rangle$, that we need to determine their evolution as well. The time evolution of these operators are given as follows,
\begin{eqnarray}\label{eq:2}
\dot{P}&=& -2\gamma_{m} P-(\omega_{m} + 2g_{2} \langle c^{\dagger} c \rangle ) X-2g_{1} \langle c^{\dagger} c \,p\rangle\non\\
		&+&2\gamma_{m}(1+2n_{th})+2 \langle p \rangle \tilde{\epsilon}_m\cos(\delta t),\non\\
\dot{X}&=& -\gamma_{m} X +2 P\omega_{m} -2(\omega_{m} + 2g_{2} \langle c^{\dagger} c\rangle ) Q-2g_{1} \langle c^{\dagger}  c q\rangle \non\\
&&+2 \langle q\rangle \tilde{\epsilon}_m\cos(\delta t).
\end{eqnarray}
The mean occupation number of phonon of a thermal reservoir is represented as $ n_{th}$. The averages of noise operators follow the vanishing statistical mean as, $\langle f(t) \rangle = 0$ where, $f$=$\xi$, $A_{in}$, $C_{in}$ and $c_{in}$. 

The system of nonlinear quantum Langevin equations (\ref{eq:1} )- (\ref{eq:2}) can be linearized by writing operators as $\langle O \rangle=\langle O_{s}\rangle+\delta \langle O \rangle$ for ($O=p, q, c, A, C, P, Q,  X$) where $O_s$ represents the steady state solution and $\delta \langle O \rangle$ is the small fluctuation around steady state. By defining the following slowly varying operators, $c\rightarrow c e^{-i\omega_{l}t}$, $A\rightarrow  A e^{-i\omega_{l}t}$ and $C\rightarrow  C e^{i(\upsilon-\omega_{l})t}$ and ignoring the fast frequency dependence of the field, the 
 steady state solution are given by $p_{s}= 0$, $q_{s}= \frac{-G_{1}c_{s}}{\omega_{m,eff}}$, $P_{s}= (1+2n_{th})$, $X_{s}= 0$, $A_{s}= \frac{-iG_a c_{s}-i\Omega C_s}{i\Delta_a+\gamma_1}$ and $C_{s}=\frac{-i\Omega A_s}{i\Delta_b +\gamma_2}$,
\begin{eqnarray}
c_{s}&=&\frac{\epsilon_{l}}{\kappa+i\Delta_{0}+[{G^2_{a}}/({i\Delta_{a}+\gamma_{1}+\frac{\Omega^2}{i\Delta_{b}+\gamma_2}})]},
\nonumber\\
Q_{s}&=& \frac{\omega_{m}P_{s}}{\omega_{m,eff}} +\frac{G_1^{2} |c_s|^2-G_1c_s\tilde{\epsilon}_m}{\omega^{2}_{m,eff}}.
\end{eqnarray} 
Here, ${\omega_{m,eff}=\omega_{m}+2g_{2}|c_{s}|^2}$ is the effective mechanical frequency. The effective linear and quadratic coupling terms are redefined as $G_1=g_1 c_s$ and $G_2=g_2 c_s$, whereas, $\Delta_{0}= \omega_{0}-\omega{l}+g_1 q_{s}+ g_2 Q_{s}$,  $\Delta_{b}= \omega_{cb} +\upsilon -\omega_{l}$ and  $\Delta_{a}= \omega_{ab}-\omega_{l}$.  
Then the corresponding linearized equations of motion are given by
  \begin{eqnarray} \label{eq:5}
\delta \dot{q}&=&\omega_{m} \delta p, \,\,\,\, \delta \dot{Q}= \omega_ m \delta \hat{X}, \non\\
\delta \dot{p}&=& -\gamma_{m}\delta p-\omega_{m,eff}\delta q-\frac{G}{2}(\delta c +\delta c^{\dagger})+\tilde{\epsilon}_m\cos(\delta t),\nonumber\\
\delta \dot{c}&=& -(\kappa +i\Delta_0)\delta c-i G_1 \delta q -iG_2 \delta Q-iG_a \delta A+\epsilon_{p}e^{-i\delta t}, \nonumber\\
\delta \dot{A}&=& -(\gamma_1+i\Delta_a) \delta A -iG_a \delta c-i\Omega \delta C, \nonumber\\
\delta \dot{P}&=&-2\gamma_{m}\delta P-\omega_{m,eff}\delta X-[G_{sm}-2\tilde{\epsilon}_m\cos(\delta t)]\delta p, \nonumber\\
\delta \dot{X}&=& 2 \omega_{m}\delta P-\gamma_m\delta X-2\omega_{m,eff}\delta Q\non\\
&& -[G_{sm}-2\tilde{\epsilon}_m\cos(\delta t)]\delta q\non\\
&&+(4G_{2}Q_{s}+2G_{1}q_{s})(\delta c+\delta c^{\dagger}),
\end{eqnarray}
where $G=2 (G_1+2 G_2 q_s)$ and $G_{sm}=2 c_s G_1$.  In above equation, the higher order terms, i.e, $\delta Q \delta c$ and $\delta q \delta c\delta c^{\dagger}$ are ignored.  In order to obtain the solution of  Eq.(\ref{eq:5}), we consider the ansatz upto first order sideband as \cite{Weis10, asjade1, asjade2}: $\delta{O}=\delta O_{+}e^{-i\delta t}+\delta O_{-}e^{i\delta t}$ where, $\delta O= (\delta p,\delta q,\delta c,\delta P,\delta Q,\delta X,\delta A,\delta C)$. By the virtue of this ansatz, we are able to obtain two sets of equations that arise due to the comparison of coefficients such as $e^{i\delta t}$ and $e^{-i\delta t}$, that are mentioned in Appendix-\ref{A}, one gets
\begin{equation}\label{c+}
\delta c_+=\frac{\epsilon_p+i \beta G_2 S}{\chi_c-\frac{i G_1 \omega_m (G-\tilde{\epsilon}_ m)}{\chi_d}+i \alpha G_2 S \omega_m+\frac{G^{2}_{a}\chi_b}{\chi_a \chi_b+\Omega^2}},
\end{equation}
	where,
\begin{eqnarray}\label{redefining}
\alpha &=&G G_{sm} \omega_m (2 \chi_d-\chi_e \chi _g)+G_t \chi_d \left(-2 \gamma^{2}_{m}+3 i \gamma_m \delta +\delta ^2\right),\nonumber\\
\beta&=&\frac{\tilde{\epsilon}_ m G_{sm} \omega^{2}_{m} (2 \chi_d-\chi_e\chi _g)}{\chi_d\chi_f (\delta +i \gamma_m)},\nonumber\\
S&=&\frac{1}{\gamma_m \chi_d \chi_f-i \delta  \chi_d\chi_f},
\end{eqnarray}
with $\chi_{a}=\gamma_1-i \delta +i \Delta_a$, $	\chi_{b}=\gamma_2-i \delta +i \Delta_b$, $\chi_{c}= -i\delta +i \Delta_{o}+\kappa$, $\chi_{d}=-i \gamma_m \delta -\delta ^2+\omega_m \omega_{m,eff}$,
	$\chi_e=2 \omega_m \omega_{m,eff}+(\gamma_m-i \delta ) (2 \gamma_m-i \delta)$,
	$\chi_{f}=2 \omega_m\omega_{m,eff} (2\gamma_m-i \delta )-i \delta \chi_e$, $\chi_g=-2 \gamma^{2}_{m}+3 i \gamma_m \delta +\delta ^2+2 \omega_m \omega_{m,eff}$ and $G_t=4 (G_1 q_s+2 G_2 Q_s)$.	

In order to study OMIT, we have to find the response of our system to the probe field of the hybrid optomechanical system. Then by using the standard input output relation which is given by $c_{out} = \sqrt{\kappa} c - c_{in} $ \cite{wall1994, Zoller2004}, the transmission of the probe field is defined as ratio of output and input cavity fields at probe frequency is given by
\begin{eqnarray}\label{transmission}
t_P &=& \frac{\epsilon_{p}-\sqrt{\kappa}\delta c_+}{\epsilon_{p}}.
\end{eqnarray}	
The quadrature $\epsilon_{out}$  the optical components in the output probe field is in general used to describe the OMIT phenomena which is given as \cite{Weis10, Safavi11}, 
\begin{eqnarray}
\epsilon_{out} &=& \frac{\sqrt{2\kappa} \delta c_{+}}{\epsilon_{p}}.
\end{eqnarray}
Here, $\nu_p$=Re$[\epsilon_{out}]$ and  $\rho_p$=Im$[\epsilon_{out}]$ correspond to the absorptive and dispersive coefficients of the output field at probe frequency. In the next section, the expression of $\delta c_+$ is used to obtain the absorptive(dispersive) profiles of the output field along with the transmission of the probe field.
\section{Results and discussion}\label{sec3}
\begin{figure}[ht]
\centering
\includegraphics[width=1\columnwidth]{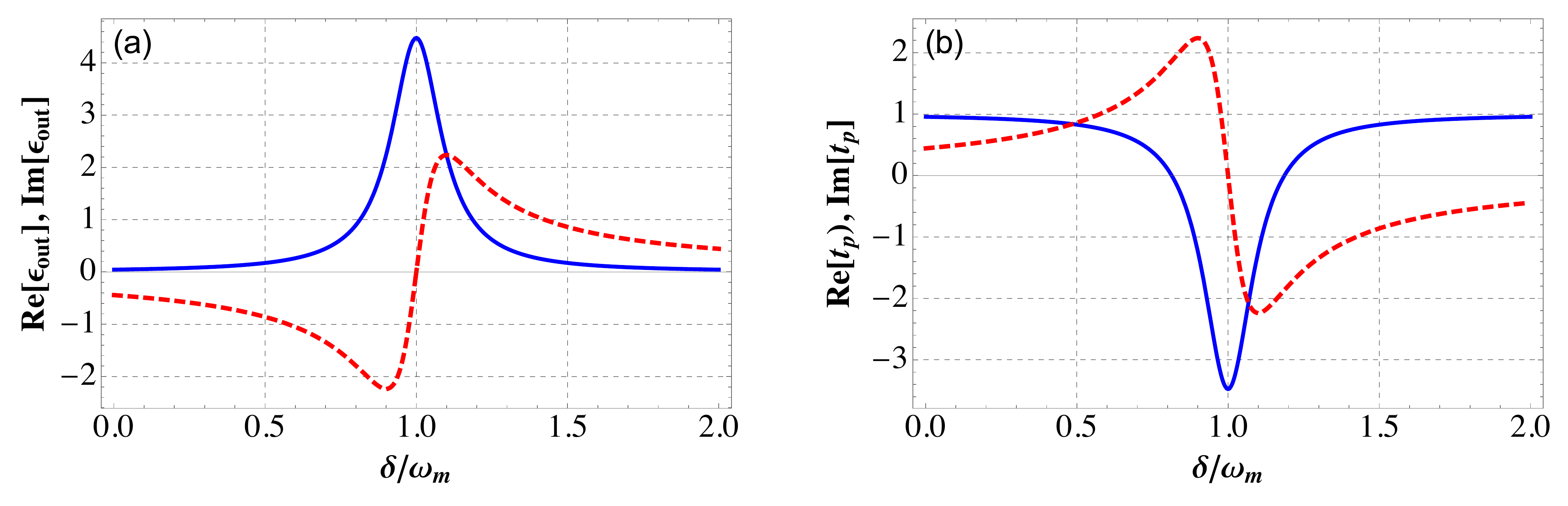}	
\caption{The blue (Red) profile show the real(imaginary) profiles of the transmission and absorption when all the coupling parameters (LOC and QOC) along with the atomic parameters are set to be zero. The selected parameters are taken as, $\Omega =0.7$, $\kappa =0.1$, $\epsilon_{p}=1$, $\hbar =1$, $\Delta_{o}=1$, $\Phi _{m}=0$, $\omega _{m,eff}=1.0006$, $\Delta_a=0.02$, $\Delta_b=0.02$, $\gamma_1=0.0001$, $\gamma _2=0.0001$, $\epsilon_l=0.05$, $\tilde{\epsilon_ m}=0$, $\omega_m=1$, $n_{th}=0$ and $\gamma_m=0.00016.$}
\label{fig:no}
\end{figure}

\begin{figure}[ht]
\centering
\includegraphics[width=1.0\columnwidth]{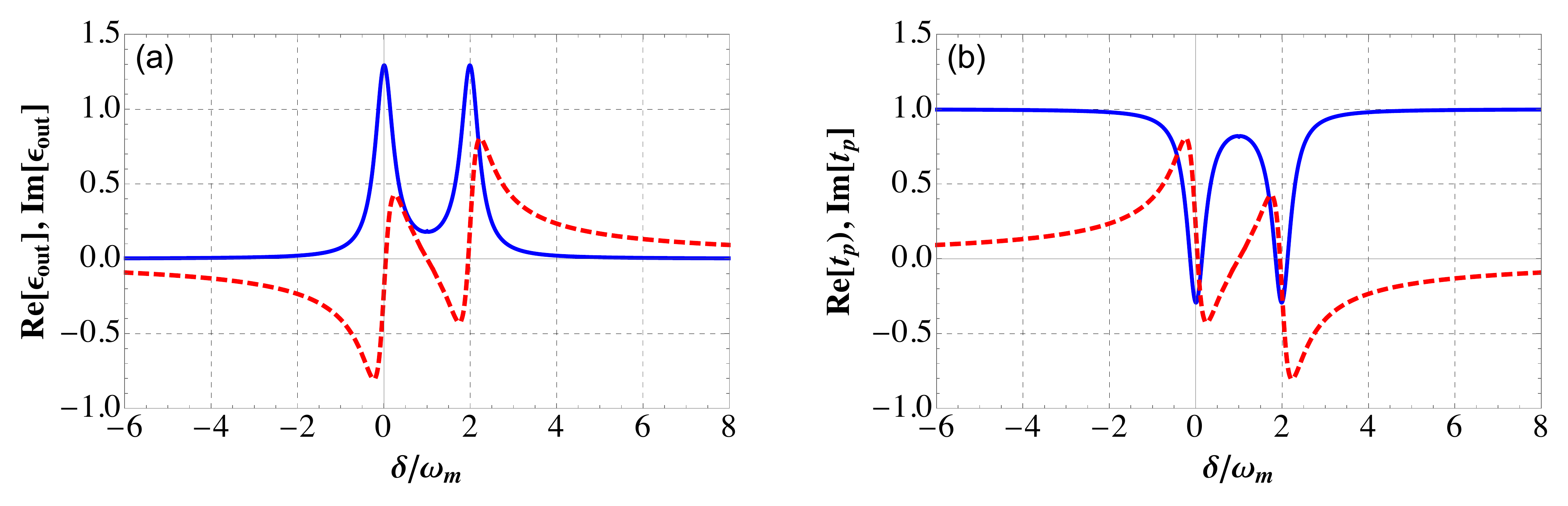}
\caption{The blue(red) curves represent the real(imaginary) profiles of both the transmission and absorption. In this plot the linear and quadratic coupling are not present whereas, the atomic parameter is present due to which a transparency window is clearly visible at $\delta=1$. The parameters are taken as, $\Omega =.01$, $\kappa =0.2$, $\epsilon_{p}=1$, $\hbar =1$, $\Delta_{o}=1$, $\Phi _{m}=0$, $\omega _{m,eff}=1.006$, $\Delta_a=1$, $\Delta_b=1$, $\gamma_1=0.30$, $\gamma _2=0.01$, $\epsilon_l=0.5$, $\tilde{\epsilon_ m}=0$, $\omega_m=1$, $n_{th}=0$, $G_{a}=1$ and $\gamma_{m}=0.001$.}
\label{fig:ga}
\end{figure}

In accordance with the generalized expression for the output probe field mentioned in Eq. (\ref{c+}) where, all the coupling parameters i.e., LOC ($G_{1}$), QOC ($G_{2}$), atomic parameter ($G_{a}$) and the driving field $\epsilon_{m}$ are present where we mainly observe the changes in the absorption(dispersion) and transmission profiles of OMIT with the variation in these aforementioned parameters. For $G_{1}=G_{2}=G_{a}=0$, a single peak is observed as shown in Fig. (\ref{fig:no}). This figure shows the absorption, dispersion and transmission profiles. The mathematical expression can be extracted from Eq. (\ref{c+}) that comes out to be as,
\begin{equation}\label{cout g1,g2,ga=0}
\epsilon_{out}=\frac{\sqrt{2\kappa} \epsilon_{p}}{-i \delta +i \Delta_o+\kappa}.
\end{equation}
From the above expression, it is clear that the denominator consists of a single root, due to which no splitting of the peak is in absorption (dispersion) and transmission profiles. The expression of transmission mentioned in Eq. (\ref{transmission}) along with Eq. (\ref{c+}) is utilized for the formation of transmission profile.
Moreover, in Fig. (\ref{fig:ga}), we observe two peaks in both the transmission and absorption (dispersion) profiles. By setting the linear optical coupling, quadratic optical coupling and the amplitude of the driving field ($\epsilon_{m}$) to zero, the expression in Eq. (\ref{c+}) reduces to the following,
\begin{equation}\label{cout,ga}
\epsilon_{out} =\frac{\sqrt{2\kappa}(\chi_{a}\chi_{b}+\Omega^{2})}{G^{2}_{a}\chi_{b}+\chi_{c}(\chi_{a}\chi_{b}+\Omega^{2})}.
\end{equation}
The above mathematical expression gives two roots that are mainly responsible for the two peaks observed in Fig. (\ref{fig:ga}). In these profiles the atomic parameter $G_a$ is taken into account whereas, linear, quadratic coupling terms between the mirror and the cavity field, along with the driving field are set to be zero. Here, the atoms also do not interact directly with the phonon or quanta of movable mirror.
\begin{figure}[h]
\centering
\includegraphics[width=1.04\columnwidth]{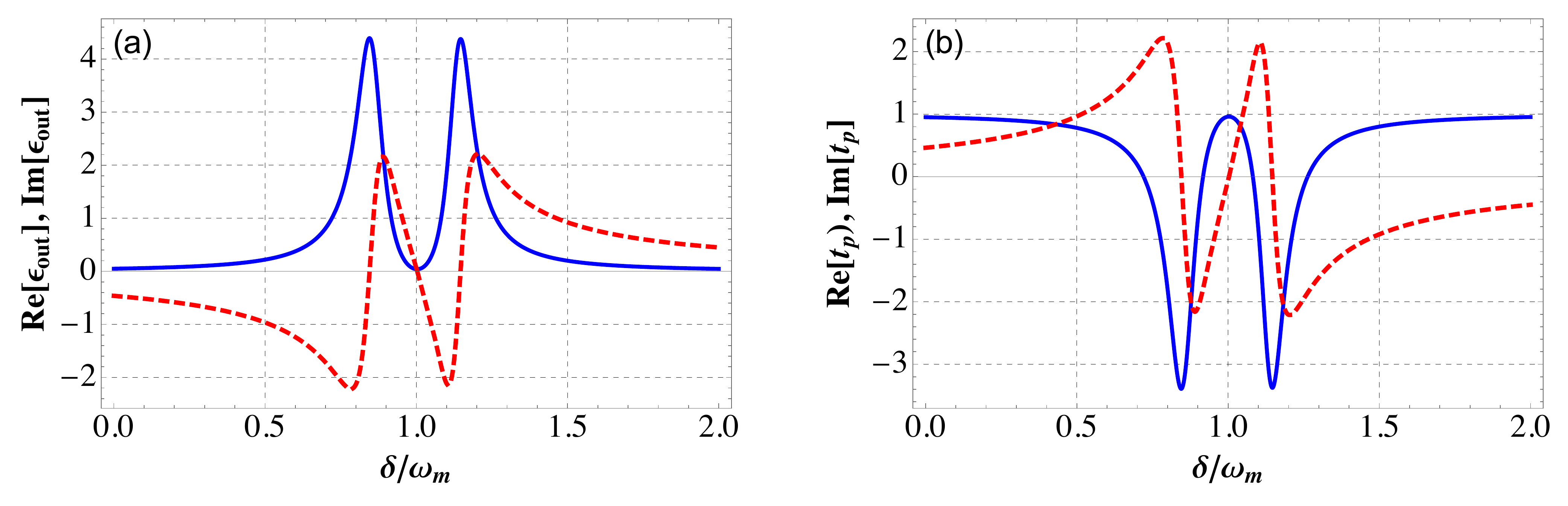}	
\caption{The blue (red) curves represent the real (imaginary) profiles of both the transmission and absorption. In this plot the atomic and quadratic coupling are not present whereas the linear coupling is present due to which a transparency window is clearly visible at $\delta=1$, $\Omega =1.0$, $\kappa=0.1$, $\epsilon_{p}=1$, $\hbar =1$, $\Delta_{o}=1$, $\Phi _{m}=0$, $\omega _{m,eff}=1.006$, $\Delta_a=1$, $\Delta_b=2$, $\gamma_1=1.01$, $\gamma _2=0.01$, $\tilde{\epsilon_ m}=0$, $\omega_m=1$, $\epsilon_l=0.5$, $n_{th}=0$, $G_{1}=0.15$ and $\gamma_m=0.004$.}
\label{fig:g1}
\end{figure}Correspondingly in Fig. (\ref{fig:g1}), we again observe two peaks in both the transmission and absorption (dispersion) profiles. These plots are again in agreement with the phenomenon of EIT. Here, the quadratic coupling and the atomic parameter are set to zero whereas, only the linear coupling is present. The mathematical expression in Eq. (\ref{c+}) now contains only the linear coupling terms, hence its reduced form is,
	\begin{equation}
		\epsilon_{out} =\frac{\sqrt{2\kappa}\chi_d}{i \tilde{\epsilon_ m} G_{1}\omega_{m}-i G G_{1}\omega_{m}+\chi_c \chi_d}.
	\end{equation}
	
\begin{figure}[h]
\centering
\includegraphics[width=1.02\columnwidth]{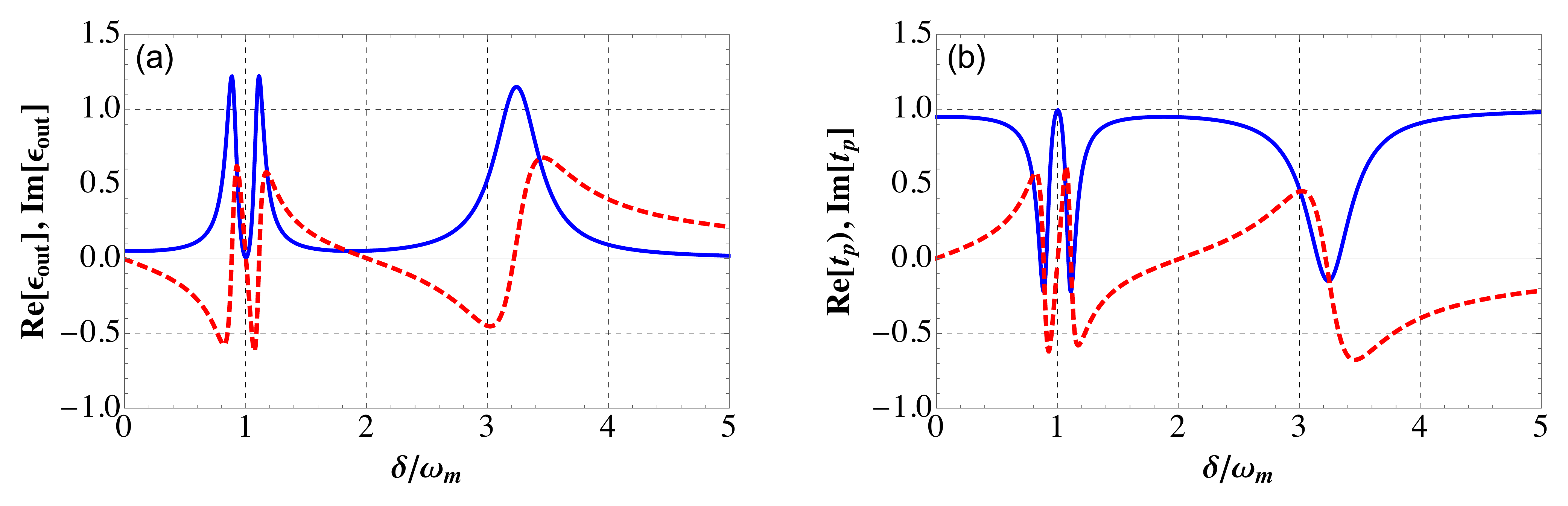}
\caption{The blue (red) curves represent the real(imaginary) profiles of both the transmission and absorption (dispersion). In this plot the linear coupling and the atomic parameter are present whereas the quadratic parameter is not present. The parameters are chosen to be as, $\Omega =1.0$, $\kappa =0.3$, $\epsilon_{p}=1$, $\hbar =1$, $\Delta_{o}=1$, $\Phi _{m}=0$, $\omega _{m,eff}=1.006$, $\Delta_{a}=1$, $\Delta_{b}=1$, $\gamma_{1}=0.35$, $\gamma_{2}=0.075$, $\epsilon_{m}=0$, $\omega_{m}=1$, $n_{th}=0$, $\epsilon_{l}=0.05$, $G_{a}=2$, $G_{1}=0.15$ and $\gamma_{m}=0.0016$.}
\label{fig:gag1}
\end{figure}

\begin{figure}[h]
\centering
\includegraphics[width=1.02\columnwidth]{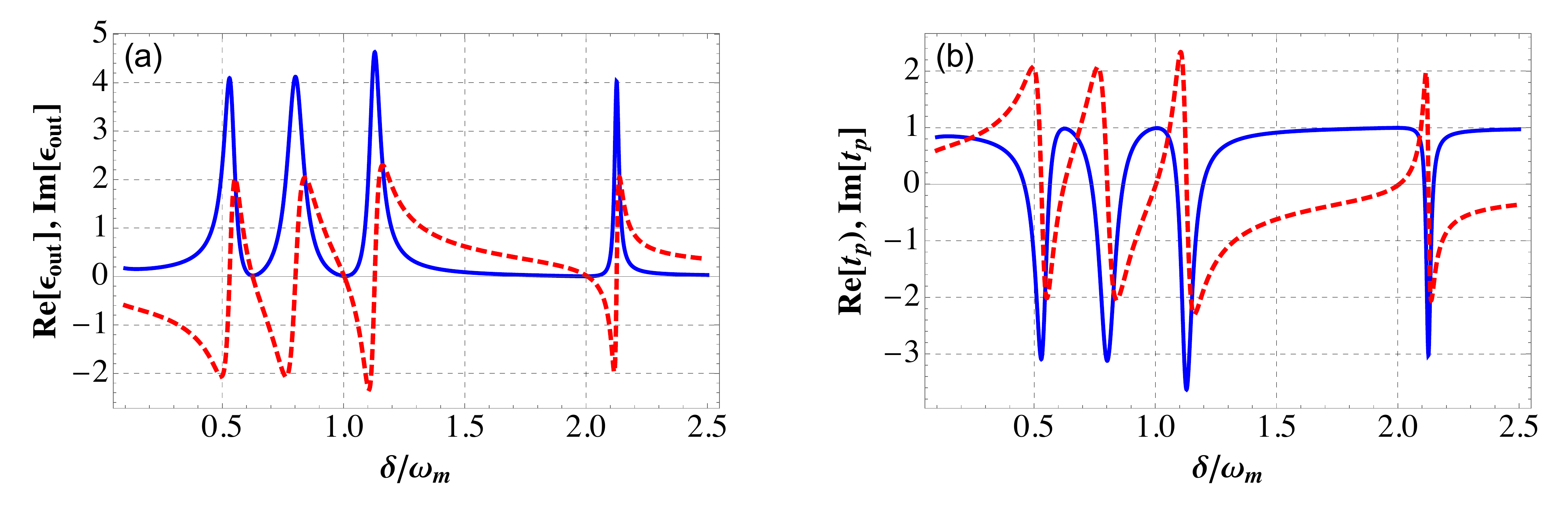}
\caption{The blue(red) curves represent the real(imaginary) profiles of both the transmission and absorption(dispersion). In this plot the linear and quadratic coupling and the atomic parameter are present simultaneously. The parameters are chosen to be as, $\Omega =0.6$, $\kappa =0.1$, $\epsilon_{p}=1$, $\hbar=1$, $\Delta_{o}=1$, $\Phi _{m}=0$, $\omega _{m,eff}=1.006$, $\Delta_{a}=0.02$, $\Delta_{b}=0.03$, $\gamma_{1}=0.001$, $\gamma _{2}=0.001$, $\epsilon_{m}=0$, $\omega_{m}=1$, $n_{th}=0$, $\epsilon_{l}=0.05$, $G_{1}=0.19$, $G_{a}=0.22$, $G_{2}=0.26$ and $\gamma_{m}=0.0015$.}
\label{fig:g1g2}
\end{figure}

\begin{figure}[h]
\includegraphics[width=0.99\columnwidth]{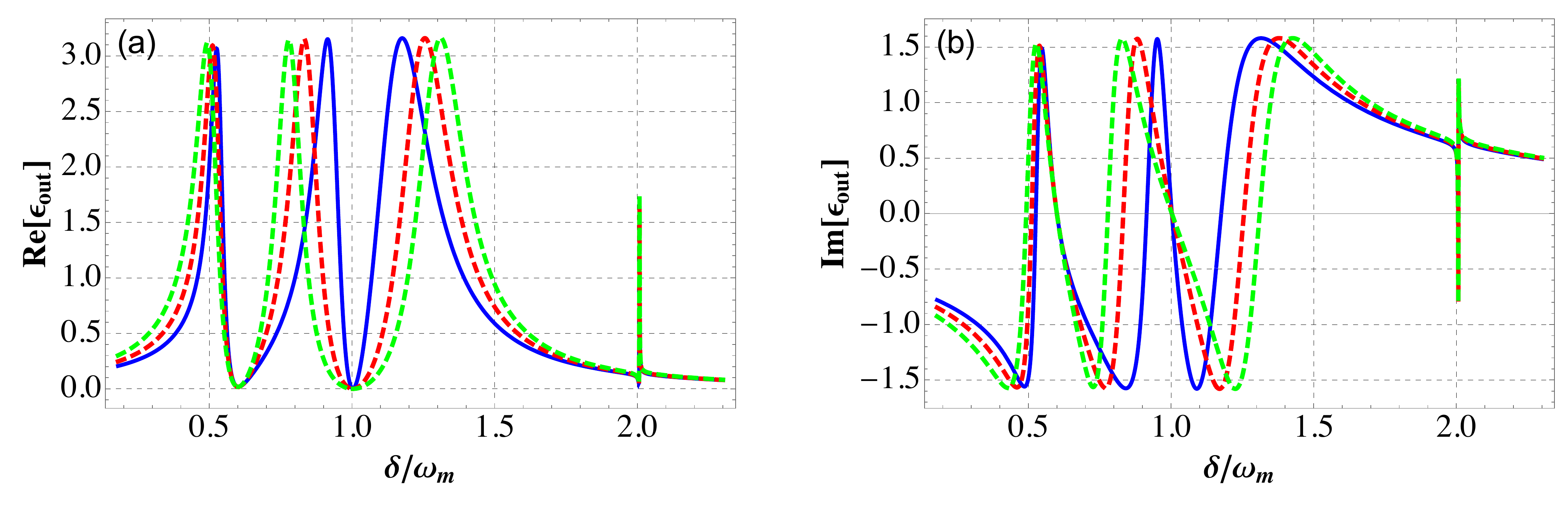}
\caption{The curves depicts the behavior of the system when the linear, quadratic and atomic parameters along with the external field are switched on. The green curve shows the shift towards the left side as well as the intensity of the QOC decreases as well when $\Phi_{m}$=$\pi$. Similarly when $\Phi_{m}$ decreases from $\pi/2$ to $0$, the peaks shift towards the right side with rise in intensity of QOC as shown in red and blue curve respectively.  $\Omega =0.5$, $\kappa =0.2$, $\epsilon_{p}=1$, $\hbar =1$, $\Delta_{o}=1$, $\Phi _{m}=0$, $\Phi _{m}=\pi$, $\Phi _{m}=\frac{\pi}{2} $, $\omega_{m,eff}=1.006$, $\Delta_{a}=0.1$, $\Delta_{b}=0.1$, $\gamma_{1}=0.001$, $\gamma_{2}=0.001$, $\epsilon_{m}=0.3$, $\omega_{m}=1$, $n_{th}=0$, $\gamma_{m}=0.00015$, $G_{1}=0.23$, $G_{a}=0.26$ and $G_{2}=0.025$.}
\label{fig:g1g2gaem}
\end{figure}

On the contrary, when both the linear coupling along with the atomic parameter are present, the absorption and transmission profiles show that two transparency windows occur at $\delta=1$ and  $\delta=3$ as depicted in Fig. (\ref{fig:gag1}). This is due to the presence of linear coupling term whereas, the second peak becomes visible, owing to the atomic parameter. When the atomic parameter $G_{a}$ is set to be zero, the second transparency window vanishes. Again, by considering Eq. (\ref{c+}), we obtain our simplified expression shown below where, the QOC and the driving field is set to zero. In this case the expression comes out to be,
	\begin{equation}
		\epsilon_{out}=\frac{\sqrt{2\kappa}\epsilon_{p} \chi_{d}  \left(\chi_{a} \chi_{b}+\Omega ^2\right)}{\chi_{b} \chi_{d}G^{2}_{a} +\left(\chi_{a}\chi_{b}+\Omega ^2\right) (\chi_{c} \chi_{d}-2iG^{2}_{1} \omega_{m})}.
	\end{equation}It is clear that three roots are obtained from the above expression that gives us three different peaks.
Subsequently, the behavior of the quadratic coupling is observed in Fig. (\ref{fig:g1g2}). The profiles distinctly show four explicit peak that arise due to the simultaneous presence of $G_a$ and both the $G_1$($G_2$) coupling terms.
Due to atomic factor we observe a dip at $\delta=0.6$ whereas, we observe two dips at $\delta=1$ and $\delta=2.2$ that arise due to the linear and quadratic coupling terms respectively that are responsible for the radiation pressure. Ideally we observe four peaks over here that arise due to four different roots of $\delta$. Keeping the side-band resolved limit i.e., $\omega_{m}>>\kappa$, the splitting between the normal modes is observed. Here, single phonon process is observed at $\delta\approx\omega_{m}$, although the external driving field of the membrane at the instant is set to zero. In such cases multiple optomechanical induced transparencies (MOMIT) are observed that are tunable by further controlling the phase and the intensity of external driving field. In tunable MOMIT, we consider the complete expression given in Eq. (\ref{c+}) together with Eq. (\ref{redefining}) by considering the linear optical coupling, quadratic optical coupling, atomic parameter and the external driving part that generates a phononic pump, due to which the variation in phase changes the intensity of the quadratic peak which is exhibited in Fig. (\ref{fig:g1g2gaem}). This comprehends us that when the phase is increased the intensity of the quadratic coupling peak decreases. Furthermore, we also observe that increasing the phase changes the origination interval of the LOC and atomic peak i.e., the peak shifts towards the left side which makes its tunable for different purposes such as slowing down the speed of the probe light as well as modification in signal propagation is quantum devices.
\section{Conclusion}\label{sec4}
In this work, we have studied the optical response of a hybrid optomechanical system and observed the optomechanical-induced transparency (OMIT) phenomenon. A three-level $\Lambda$-type atomic ensemble placed inside an optical cavity is considered. We have shown that multiple transparency windows can be attained owing to the simultaneous presence of linear optical coupling (LOC), quadratic optical coupling (QOC), and atomic parameters. The strong coupling of the dielectric membrane with the optical cavity is the optimal condition to obtain quantum control of the mechanical system. It then follows that the induced transparency windows are optimized by controlling the LOC and the QOC. The constructive and destructive interferences can be also adjusted by modifying the external driving fields.
\appendix
\section{Equation of motions}
\label{A}
The set of equations utilizes to acquire the final expression of $\delta c_{+}$ in Eq. (\ref{c+}) is given by
\begin{eqnarray}\label{eq:7}
-i\delta(\delta q_{+})&=&\delta p_{+}\omega_ m \\
		(-i\delta +\gamma_m) (\delta p_{+})&=&-\omega_{m, eff}\delta q_{+}-G\delta c_{+}+\tilde{\epsilon}_m,\\
		\chi_{c}(\delta c_+)&=&  -i(G_{1}\delta q_{+}+G_{2}\delta Q_{+}) +\epsilon_p \non\\
		&-&iG_{a}\delta A_+,\\
		\chi_{a}(\delta A_{+})&=&-iG_{a}\delta c_+-i\Omega \delta C_+,\\
	\chi_{b} \delta C_{+}&=& -i\Omega \delta A_+,\\
		-i\delta (\delta Q_{+})&=&\omega_m \delta X_+,\\
		(-i\delta +2\gamma_m )(\delta P_{+})&=& -\omega_{m,eff}\delta X_+-G_{sm} \delta p_+,\\
		(-i\delta +\gamma_m) (\delta X_{+})&=& 2 \omega_{m}\delta P_+-2\omega_{m,eff}\delta Q_+  \nonumber\\
		&-&G_t \delta c_+-G_{sm}\delta q_{+},
	\end{eqnarray}
where $\chi_{a_o}=\gamma_1+i \delta +i \Delta_a$, $	\chi_{b_o}=\gamma_2+i \delta +i \Delta_b$, $\chi_{c_o}= i\delta +i \Delta_{o}+\kappa$. Similarly the equations for $\delta O_{-}$ can be collected as follows. 
	\begin{eqnarray}
		i\delta (\delta q_{-})&=& \delta p_{-}\omega_ m,\\
       (i\delta +\gamma_m)(\delta p_{-})&=&-\omega_{m, eff} \delta q_{-}-G\delta c_-+\tilde{\epsilon}_m,\\
		\chi_{c_o}(\delta c_{-})&=&  -i(G_{1}\delta q_{-}+G_{2}\delta  Q_{-}) \non \\
		&-&		iG_{a}\delta A_{-},\\
		\chi_{a_o}(\delta A_{-})= && -iG_{a}\delta c_{-} -i\Omega \delta C_{-},\\
		\chi_{b_o}(\delta C_{-})&=& -i\Omega   \delta A_{-},\\
		i\delta (\delta Q_{-})&=&\omega_m \delta X_{-},\\
		(i\delta +2\gamma_m )(\delta P_{-})&=& -\omega_{m,eff}\delta X_{-}-G_{sm}\delta p_{-},\\
		(i\delta +\gamma_m) (\delta X_{-})&=& 2 \omega_{m}\delta P_{-}-2\omega_{m,eff}\delta Q_{-}\non \\
		&-& G_t \delta c_- -G_{sm}\delta q_{-}.
	\end{eqnarray}


\begin{thebibliography}{100}
\bibitem{Harris} S. E. Harris, Phys. Today {\bf 50}, 36 (1997).	
\bibitem{zubairy} M. O. Scully and M. S. Zubairy, \emph {Quantum Optics, Cambridge University Press} (1997).
		\bibitem{hau1999} L. V. Hau, S. E. Harris, Z. Dutton and C. H. Behroozi, Nat. {\bf397}, 594–598 (1999).
		\bibitem{scully1999} M. M. Kash, V. A. Sautenkov, A. S. Zibrov, L. Hollberg, G. R. Welch, M. D. Lukin, Y. Rostovtsev, E. S. Fry, and M. O. Scully, Phys. Rev. Lett. {\bf82}, 5229 (1999).
\bibitem{Boutabba}  N Boutabba, H Eleuch, Applied Mathematics \& Information Sciences {\bf7}, 1505 (2013).
		\bibitem{maier2016}R. Lechner, C. Maier, C. Hempel, P. Jurcevic, B. P. Lanyon, T. Monz, M. Brownnutt, R. Blatt, and C. F. Roos, Phys. Rev. A {\bf93}, 053401 (2016).
		\bibitem{Hakuta1993}G. Z. Zhang, K. Hakuta, and B. P. Stoicheff, Phys. Rev. Lett. {\bf71}, 3099 (1993).
		\bibitem{Hemmer1995}P. R. Hemmer, D. P. Katz, J. Donoghue, M. Cronin-Golomb, M. S. Shahriar, and P. Kumar, Opt. Lett. {\bf20}, 982–984 (1995).
		\bibitem{Agarwal2010}S. Huang and G. S. Agarwal, Phys. Rev. A {\bf81}, 033830 (2010).
		\bibitem{R2010}S. Weis, R. Riviere. S. Deleglise, E. Gavartin, O. Aecizet, A. Schliesser, and T. J. kippenberg, Sci. {\bf330}, 1520-1523 (2010).
		\bibitem{Safavi2011} A. H. Safavi-Naenin, T. P. Mayer Alegre, J. Chan, M. Eichenfield, M. Winger, Q. Lin, J. T. Hill, D. E. Chang, and O. Painter, Nat. {\bf472}, 69 (2011).
		\bibitem{singh2014}V. Singh, S. J. Bosman, B. H. Schneider, Y. M. Blanter, A. Castellous-Gomez, and G. A. Steele, Nat. Nanotechnol. {\bf9}, 820 (2014).
		\bibitem{Agarwal2013} K. N. Qu, and G. S. Agarawal, Phys. Rev. A {\bf87}, 031802 (2013).
		\bibitem{zhou2012} F. Hocke, X. Zhou. A. Schliesser, T. J. Kippenberg, H. Huebl, and R. Gross, New. J. Phys. {\bf14}, 123037 (2012).
		\bibitem{Wu2007} Y. Wu and X. Yang, Phys. Rev. A {\bf76}, 013832 (2007).
		\bibitem{X2011} W. -X. Yang, A. -X. Chen, R.-K. lee, and Y. Wu, Phys. Rev. A {\bf84}, 013835 (2011).
		\bibitem{X2010} X. -T. Xie and M. A. Macovei, Phys. Rev. Lett. {\bf104}, 073902 (2010).
		\bibitem{Rabl2009}P. Rabl, P. Cappellaro, M. V. Gurudev Dutt, L. Jiang, J. R. Maze, and M. D. Lukin, Phys. Rev. B {\bf79}, 041302(R) (2009).
		\bibitem{Michael2007} M. Eichenfield, C. P. Michael, R. Perahia, and O. Painter, Nat. Photonics {\bf1}, 416–422 (2007).
		\bibitem{xiong2008} Mo Li, W. H. P. Pernice, C. Xiong, T. Baehr-Jones, M. Hochberg and H. X. Tang, Nat. {\bf456}, 480–484 (2008).
		\bibitem{Sete2015}E. A. Sete, H. Eleuch and C. H. R. Ooi, Phys. Rev. A {\bf92}, 033843 (2015).
		\bibitem{Eleuch2012}E. A. Sete and H. Eleuch, Phys. Rev. A {\bf85}, 043824 (2012).
		\bibitem{Juuso} J. Manninen, M. Asjad, E. Selenius, R. Ojajarvi, P. Kuusela and F. Massel, Phys. Rev. A {\bf 98}, 043831 (2018).
		\bibitem{Aspelmeyer14}  M. Aspelmeyer, T. J. Kippenberg and F. Marquardt, Rev. Mod. Phys. {\bf 86}, 1391 (2014).
		\bibitem{qoptm}W. P. Bowen and G. J. Milburn,\emph{Taylor and Francis Group, LLC} (2016).
		\bibitem{Eleuch2015}E. A. Sete and H. Eleuch, Phys. Rev. A {\bf91}, 032309 (2015).
		\bibitem{chang2011}Y. Chang, T. Shi, Yu-xi. Liu, C. P. Sun, and F. Nori, Phys. Rev. A {\bf83}, 063826 (2011).
		\bibitem{han2011}Y. Han, J. Cheng, and L. Zhou, J. Phys. B: At. Mol. Opt. Phys. {\bf44}, 165505 (2011).
		\bibitem{kronwald2013} A. Kronwald and F. Marquardt, Phys. Rev. Lett. {\bf111}, 133601 (2013).
		\bibitem{Borkje2013} K. Borkje, A. Nunnenkamp, J. D. Teufel and S. M. Girvin, Phys. Rev. Lett. {\bf111}, 053602 (2013).
		\bibitem{H2012} H. Xiong, L. G. Si, A. S. Zheng, X. Yang, and Y. Wu, Phys. Rev. A {\bf86}, 013815 (2012).
		\bibitem{H2016} H. Xiong, L. G. Si, X. Y. Lu and Y. Wu, Opt. Express {\bf24}, 5773 (2016).
		\bibitem{Sankey2010} J. C. Sankey, C. Yang, B. M. Zwickl, A. M. Jayich and J. G. E. Harris, Nat. Phys. {\bf6}, 707 (2010).
		\bibitem{asjad2013}   M. Asjad and D. Vitali, J. Phys. B: At. Mol. Opt. Phys.  {\bf47}, 045502 (2013).
		\bibitem{asjad2014}  M. Asjad, G. S. Agarwal, M. S. Kim, P. Tombesi, G. Di Giuseppe and D. Vitali, Phys. Rev. A  {\bf89}, 023849 (2014).
		\bibitem{Murch2008} K. W. Murch, K. L. Moore, S. Gupta and D. M. Stamper-Kurn, Nat. Phys. {\bf4}, 561 (2008).
		\bibitem{Purdy2010} T. P. Purdy, D. Brooks, T. Botter, N. Brahms, Z.-Y. Ma and D. M. Stamper-Kurn, Phys. Rev. Lett. {\bf105}, 133602 (2010).
		\bibitem{Karuza2013} M. Karuza, C. Biancofiore, M. Bawaj, C. Molinelli, M. Galassi. R. Natali, P. Tombesi, G. Di. Giuseppe, and D. Vitali, Phys. Rev. A {\bf88}, 013804 (2013).
		\bibitem{Zhan2013} X. G. Zhan, L. G. Si, A. S. Zheng, and X. Yang, J. Phys. B {\bf46}, 0255021 (2013).
		\bibitem{Rugar2004} D. Rugar, R. Budakian, H. J. Mamin and B. W. Chui, Nat. {\bf430}, 329 (2004).
		 \bibitem{Candeloro} A. Candeloro, S. Razavian, M. Piccolini, B. Teklu, S. Olivares, and M. G. A. Paris, Entropy {\bf23}, 1353 (2021).
		\bibitem{Braginsky2002} V. Braginsky and S. P. Vyatchanin, Phys. Lett. A {\bf293}, 228 (2002).
		\bibitem{Vitali07}  D. Vitali, S. Gigan, A. Ferreira, H. R. Bohm, P. Tombesi, A. Guerreiro, V. Vedral, A. Zeilinger and M. Aspelmeyer, Phys. Rev. Lett. {\bf 98}, 030405 (2007).
		\bibitem{Korppi18} C. F. Ockeloen-Korppi, E. Damskagg, J. M. Pirkkalainen, M. Asjad, A. A. Clerk, F. Massel, M. J. Woolley and M. A. Sillanpaa, Nat. {\bf 556}, 478 (2018).
		\bibitem{asjad15} M. Asjad, S. Zippilli, P. Tombesi and D. Vitali, Phys. Scr. {\bf 90}, 074055 (2015).
		\bibitem{asjad16} M. Asjad, S. Zippilli and D. Vitali, Phys. Rev. A {\bf 93}, 062307 (2016).
		\bibitem{Berihu} B. Teklu, T. Byrnes, and F. S. Khan, Phys. Rev. A  {\bf97}, 023829 (2018).
		
\bibitem{Feizpour} A. Feizpour, G. Dmochowski, and A. M. Steinberg, Phys. Rev. A {\bf93}, 013834 (2016).


\bibitem{Weis10} S. Weis, R. Riviere, S. Deleglise, E. Gavartin, O. Arcizet, A. Schliesser and T. J. Kippenberg, Sci. \textbf{330}, 1520 (2010).
\bibitem{asjade1} M. Asjad, J. Russ. Laser Res. \textbf{34}, 278 (2013).
\bibitem{asjade2} M. Asjad, J. Russ. Laser Res. \textbf{34}, 159 (2013).
\bibitem{wall1994} D. F. Walls and G. J. Milburn, \emph{Quantum Optics Springer-Verlag, Berlin Heidelberg} (1994).
\bibitem{Zoller2004}C. W. Gardiner, P. Zoller, \emph{Quantum Noise, Springer Berlin Heidelberg} (2004).
\bibitem{Safavi11} A. H. Safavi-Naeini, T. P. Mayer Alegre, J. Chan, M. Eichenfield, M. Winger, Q. Lin, J. T. Hill, D. E. Chang and O. Painter, Nat. \textbf{472}, 69 (2011).
\bibitem{sing2022}S. K. Singh, M. Parvez, T. Abbas, J. -Xin. Peng, M. Mazaheri and M. Asjad. Phys. Lett. A {\bf442}, 128181 (2022).
	\end{thebibliography}
\end{document}